\documentclass[aps,twocolumn,amsmath,amssymb,showpacs,showkeys,floatfix]{revtex4}
\usepackage{graphicx}
\usepackage{amssymb}
\usepackage{color}
\usepackage{float}
\usepackage{mathrsfs}

\begin{document}



\title{Dipole response in neutron-rich nuclei with new Skyrme interactions}

\author{H. Zheng$^{1}$\footnote{Email address: zheng@lns.infn.it}, S. Burrello$^{1,2}$,  M. Colonna$^{1}$, V. Baran$^{3}$}
\affiliation{$^{1}$ Laboratori Nazionali del Sud, INFN, I-95123 Catania, Italy}

\affiliation{$^{2}$ Physics and Astronomy Department, University of Catania, Italy}
\affiliation{$^{3}$ Faculty of Physics, University of Bucharest, Romania}


\begin{abstract}
We investigate the isoscalar and isovector
E1 response of neutron-rich nuclei, within a semi-classical transport
model employing effective interactions for the nuclear mean-field. 
In particular, we adopt the recently introduced
SAMi-J Skyrme interactions, whose parameters are specifically tuned to
improve the description of spin-isospin properties of nuclei.
Our analysis evidences a relevant degree of isoscalar/isovector mixing
of the collective excitations developing in neutron-rich systems. 
Focusing on the low-lying strength emerging in the isovector response, 
we show that
this energy region essentially corresponds to the excitation of isoscalar-like
modes, which also contribute to the isovector response owing to their
mixed character.     
Considering effective interactions which mostly differ in the isovector channels, 
we observe that these  mixing effects increase with the slope $L$ of the symmetry energy at saturation density, leading to a larger strength in the low-energy region of the isovector
response.  This result appears connected to the increase, with $L$,  
of the neutron/proton asymmetry at
the surface of the considered nuclei, i.e., to the extension of the 
neutron skin.

\end{abstract}



\maketitle


\section{Introduction}

The dynamics of many-body interacting systems often manifests the development of collective
patterns.  The understanding of such fascinating properties of complex systems 
is quite helpful to shed light on fundamental properties of the interaction among the constituent
particles. 
In particular, the investigation of collective phenomena represents one of the 
most challenging and lively research fields in nuclear physics. 
In nuclei, giant resonances are well established
collective states, with an energy larger than the particle
separation energy  \cite{Har2001}.   
An example is the giant dipole resonance (GDR), which is still the object 
of intense investigation.   
With the advent of
the first-generation exotic-beam facilities, much attention was directed towards the features 
of the  collective (multipole)  response  of  unstable
nuclei.  Restricting the discussion to isovector dipole excitations of
neutron-rich systems, one generally observes 
a stronger fragmentation of strength than in
nuclei with small neutron excess,  with  significant  components  located  in  an
energy  domain  well  below  that  of  the GDR
\cite{harPRL2000,harPRC2002,konPRC2012,adrPRL2005,kliPRC2007,carPRC2010,wiePPNP2011,tamPRL2011,ni3,sn2}.
The  nature  of  the  low-lying  excitations  is  still a  matter  of
ongoing  discussion
\cite{aumPS2013,savPPNP2013,paaRPP2007,lanPRC2011,Crespi2014,Repko2013}.
Unlike the GDR, where neutrons
and protons move against each other, 
this low-lying strength could be associated with 
an oscillation of the outermost 
neutrons 
(neutron skin) against the
N=Z core. This mode is commonly  referred  to  as  a `soft'  or
`pygmy' dipole resonance (PDR). This interpretation was
already discussed in the early 1990s \cite{suz1990,isaPRC1992} and is strongly
supported  by  recent  relativistic random phase approximation (RPA) calculations  \cite{tsoPLB2004,paar2003,vretenar2001}.
On the other hand, 
other microscopic studies predict 
a larger fragmentation of the GDR strength  \cite{mazPRC2012}  and
the absence of collective states in the low-lying excitation region \cite{sarPLB2004},
thus relating the observed strength to a particular structure of the single-particle levels.
Therefore  a number of critical
questions concerning the nature of the PDR still remain. 


It is worth noting that 
the low-lying electric dipole
E1 strength  in  unstable  neutron-rich  nuclei  is
currently discussed also in the astrophysical context,
in connection with the reaction rates
in the r-process nucleosynthesis. 
It appears that the existence of the pygmy mode 
could have a strong impact on the abundances of the elements
in the Universe \cite{gor2004}. 
Moreover, as it has been evidenced in mean-field based calculations, 
the features of neutron-rich nuclei, such as
pygmy mode and neutron skin, are clearly related to the isovector 
terms of the nuclear effective interactions (or modern Energy Density Functional theories). 
These terms are linked to the symmetry energy contribution in the nuclear
Equation of State (EoS), a concept which is widely employed 
in the description of heavy ion collisions \cite{barPR2005,EPJA,Hua} and
also
in astrophysics, as far as the modelization of supernova
explosion and neutron stars is concerned \cite{stePR2005,baoPR2008,burrelloPRC2015}.


The aim of the present
paper is therefore to investigate the dipole response of neutron-rich nuclei, by  
solving the semi-classical Vlasov equation.
In the past, studies based on semi-classical approaches, such as the 
Goldhaber-Teller (GT) \cite{goldhaber1948}
or the Steinwedel-Jensen (SJ) \cite{stein1950} models, 
have given an important contribution to the understanding of the
main features of giant resonances and of their link to important
nuclear properties, such as compressibility and symmetry energy. 
In particular,  the Vlasov equation has already been shown to describe reasonably 
well some relevant properties of different collective excitations of nuclei
\cite{Brink1986,Burgio1988,Matera}. 
It is clear that, within such a semi-classical description,
shell effects, certainly important in shaping the fine structure
of the dipole response \cite{mazPRC2012},  are absent.
However, the genuine collective features of the nuclear excitations should naturally come out from this analysis.

Here, for the mean-field representation,  we will employ new effective
interactions, of the Skyrme type - the SAMi-J interactions - which have been especially devised to
improve the description of spin-isospin properties of nuclei \cite{coll4}.
We will focus on the mixed isoscalar/isovector character of the collective 
excitations in neutron-rich nuclei, in some analogy with features already discussed in the context of infinite nuclear matter, where the degree of mixing is observed to increase with the isospin asymmetry, tuned
by the density dependence of the symmetry energy \cite{barPR2005,EPJA}.    
Then, we show that the relative isoscalar/isovector weight of the different modes, as observed in the nuclear response, is  determined by  their intrinsic structure,
in terms of isoscalar (IS) and isovector (IV) components,  as well as by the type of initial perturbation considered.
As a result, within our framework, the low-lying strength arising in the IV dipole response essentially reflects the partial isovector character of collective modes which are mostly isoscalar-like, in agreement with previous 
semi-classical and RPA studies \cite{urbPRC2012,mazPRC2012}.  

An important goal of our investigation is to get a deeper insight into the link
between the nuclear response and the properties of the underlying effective interaction.
In particular, considering SAMi-J parametrizations which mostly 
differ in the isovector channel, we will explore the relation
between the
mixed isoscalar/isovector structure of the dipole collective modes and 
the density dependence of the symmetry energy.  We notice that the latter quantity also affects 
the size of the neutron skin.
Thus our analysis also aims at elucidating the 
possible connection 
between the strength observed, for selected nuclei,  in the PDR region and the corresponding
neutron skin extension \cite{coll2}. 

The paper is organized as follows: in Section II we outline the theoretical framework and the 
main ingredients associated with
the Vlasov equation and its numerical solution.  The different Skyrme parametrizations employed 
in our study are presented. 
The results concerning the isoscalar and isovector dipole response, for 
selected nuclei in three
different mass regions are discussed in Section III.A. Two different kinds of initial perturbation, corresponding to standard isoscalar and isovector excitations, are considered. 
The corresponding transition densities are presented in Section III.B.
Finally, in Section IV, conclusions and perspectives are drawn.

\section{Theoretical framework}

The Vlasov equation, which describes the time evolution of the one-body distribution function 
in phase space, represents the semi-classical limit of Time-Dependent Hartree-Fock (TDHF)
and, for small-oscillations, of the RPA equations.
While the model is unable to account for effects associated with the shell structure,
this self-consistent approach is suitable to describe robust quantum modes, of zero-sound type,
in both nuclear matter and finite nuclei \cite{barPR2005,Burgio1988,urbPRC2012,barPRC2012}. 
One has essentially to solve the two coupled Vlasov kinetic equations for the neutron and proton
 distribution functions $f_q({\bf r},{\bf p},t)$, with $q=n,p$ 
\cite{barPR2005}:
\begin{equation}
\frac{\partial f_q}{\partial t}+\frac{\partial \epsilon_q}{\partial {\bf p}}\frac{\partial f_q}{\partial {\bf r}}-
\frac{\partial \epsilon_q}{\partial {\bf r}}\frac{\partial f_q}{\partial {\bf p}}=0. 
\label{vlasov}
\end{equation}
In the equations above, $\epsilon_q$ represents the single particle energy, which can be deduced from the
energy density, $\mathscr{E}$. 
Considering a standard Skyrme interaction, the latter is expressed in terms of the isoscalar, $\rho=\rho_n+\rho_p$,
and isovector, $\rho_{3}=\rho_n-\rho_p$,  densities and 
kinetic energy densities ($\tau=\tau_{n}+\tau_{p}, \tau_{3}=\tau_{n}-\tau_{p}$) as \cite{radutaEJPA2014}:
\begin{eqnarray}
\mathscr{E}&=&\frac{\hbar^2}{2 m}\tau + C_0\rho^2 + D_0\rho_{3}^2 + C_3\rho^{\alpha + 2} + D_3\rho^{\alpha}\rho_{3}^2 ~+ C_{eff}\rho\tau \nonumber\\
&& + D_{eff}\rho_{3}\tau_{3} + C_{surf}(\bigtriangledown\rho)^2 + D_{surf}(\bigtriangledown\rho_3)^2,
\label{eq:rhoE}
\end{eqnarray}
where $m$ is the nucleon mass and the coefficients $C_{..}$, $D_{..}$ are combinations of traditional Skyrme parameters. 
In particular, the terms with coefficients $C_{eff}$ and $D_{eff}$ are the momentum dependent contributions to the nuclear 
effective interaction.  The Coulomb interaction is also considered in the calculations.  
We are mostly interested in the effects linked to the isovector terms, thus we introduce 
the symmetry energy per nucleon, $E_{sym}/ A = C(\rho) I^2$, 
where $I = (N-Z)/A$ is the asymmetry parameter and the coefficient $C(\rho)$ can be written as a function of the Skyrme coefficients (at temperature T = 0):
\begin{equation}
C(\rho) = \frac{\epsilon_F}{3} + D_0\rho + D_3\rho^{\alpha+1} ~+ 
\frac{2m}{\hbar^2}\left(\frac{C_{eff}}{3} + D_{eff}\right)\epsilon_F\rho,
\end{equation}
with $\epsilon_F$ denoting the Fermi energy at density $\rho$. 

\begin{figure}
\includegraphics[scale=0.4]{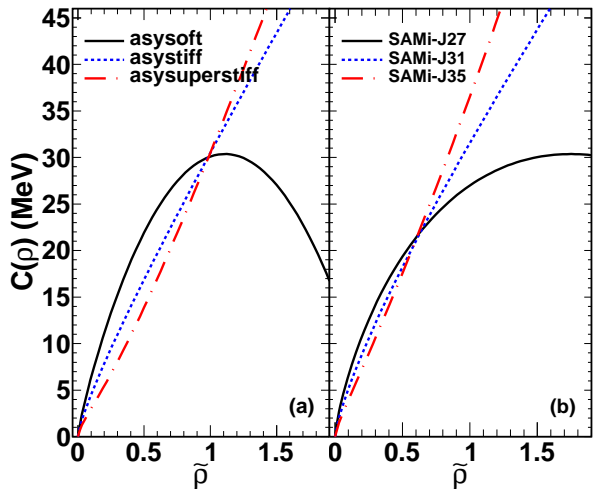}
\caption{(Color online) The symmetry energy versus reduced density $\tilde\rho=\frac{\rho}{\rho_0}$ for the EoS 
without (panel (a)) and with (panel (b))
momentum dependent terms. 
}
\label{fig01} 
\end{figure}

In the following we will adopt the recently introduced SAMi-J Skyrme effective interactions \cite{coll4}. 
The corresponding parameters have been fitted based on the SAMi fitting protocol \cite{coll4}: 
binding energies and charge radii of some doubly magic nuclei which allow the SAMi-J family to 
predict a reasonable saturation density ($\rho_0=0.159$ fm$^{-3}$), 
energy ($E/A (\rho = \rho_0) =-15.9$ MeV) and incompressibility ($K = 245$ MeV) of symmetric  nuclear  matter;
some selected spin-orbit splittings; the spin and spin-isospin sensitive Landau Migdal parameters
\cite{LMparam}; 
and, finally, the neutron matter EoS of Ref.\cite{wir19}. 
These  features allow the new SAMi interactions to give a reasonable description of 
isospin as well as  spin-isospin  resonances, keeping a good reproduction of well 
know empirical data such as masses, radii and important nuclear excitations. 
The main difference between SAMi and the SAMi-J family is that SAMi-J has been produced by 
systematically varying the value of $J = C(\rho_0)$ from 27 to 35 MeV, keeping fixed the optimal value of 
the incompressibility and effective mass predicted by SAMi and refitting again the parameters 
for each value of J. This produces a set of interactions of similar quality on the isoscalar 
channel and that, approximately,  isolate the effects of modifying the isovector channel in the study 
of a given observable.
In our calculations,
we employed, in particular, three SAMi-J parametrizations: SAMi-J27, SAMi-J31 and SAMi-J35 \cite{coll4}. 
Since, as mentioned above, the SAMi-J interactions have been fitted in order to also reproduce the main features of finite nuclei, 
for the three parametrizations the symmetry energy coefficient gets the same value, $C(\rho_c) \approx 22$ MeV at $\rho_c = 0.65\rho_0$, which can be taken as the average density of medium-size nuclei. 
Thus the curves representing the density dependence of $C(\rho)$ cross each other at $\rho = \rho_c$, 
i.e., below saturation density, see Fig.\ref{fig01} (panel (b)).
The corresponding values of symmetry energy at saturation, together with the values of the slope parameter
 $\displaystyle L = 3 \left. \rho_0 \frac{d C(\rho)}{d \rho} \right\vert_{\rho=\rho_0}$ are reported in Table I. 
In the following we will also indicate the SAMi-J interactions as
momentum dependent (MD) interactions.

In order to make a connection with previous studies, we shall also consider simplified Skyrme interactions  where the momentum dependent terms
are neglected  ($C_{eff} = D_{eff} = 0$), corresponding to an incompressibility modulus equal to $K = 200$ MeV \cite{coll2}.  
We will refer to these interactions as momentum-independent
(MI) interactions. 

As far as the symmetry energy is concerned, 
the parametrizations considered 
allow for three different types of density dependence, 
associated with three different parametrizations of the potential part of the symmetry energy coefficient, $C_{pot}(\rho)$. 
For the {\it asystiff} EoS, $\displaystyle C_{pot}(\rho) = 18 \rho/\rho_0$ MeV. 
The {\it asysoft} case corresponds to a SKM* Skyrme-like parametrization 
with $\displaystyle {C_{pot}(\rho)} = 0.5 \rho(482-1638 \rho)$ MeV, associated with a small value 
of the slope parameter $L$.  
Lastly, for the {\it asysuperstiff} EoS, $C_{pot}(\rho)=18\frac{\rho}{\rho_0} \frac{2 \rho}{(\rho + \rho_0)}$ MeV, the symmetry term increases rapidly around saturation density, being characterized by a large value of the slope parameter.
The corresponding values are listed in Table I.  
As one can see from Fig.\ref{fig01} (panel (a)), the three parametrizations of the symmetry energy cross each other at 
$\rho = \rho_0$ in this case.


The integration of the transport equations is based on the test-particle (t.p.) (or pseudo-particle) method \cite{wong},
with a number of $1500$ t.p. per nucleon in all the cases, ensuring in this way a good spanning of the phase space. 
In order to determine the ground state configuration of the nuclei under study, 
one should find the stationary solution of Eq.(\ref{vlasov}).  We adopt the following numerical procedure:  
neutrons and protons are distributed inside spheres of 
radii $R_n$ and $R_p$, respectively. Accordingly, particle momenta are initialized inside Fermi spheres associated
with the local neutron or proton densities.  Then   $R_n$ and $R_p$ are tuned in order to minimize the corresponding total energy,
associated with the effective interaction adopted in the calculations. 
We note here that the test particle method
is able to reproduce accurately the equation of state of nuclear matter and provide reliable results regarding the properties of nuclear surface \cite{idiNPA1993} and ground state energy for finite nuclei \cite{schPPNP1989,coll2}.

From the one-body distribution functions one obtains the local densities: 
\begin{equation}
\rho_q({\bf r},t)=\frac{2}{(2\pi\hbar)^3}\int d^3p f_q({\bf r},{\bf p},t),
\end{equation}
as well as the average value of the radial distance ${\bf r}$  to the power $n$:
\begin{equation}
\displaystyle \langle{\bf r}_q^n \rangle = \frac{1}{N_q} \int d^3r~{\bf r}^n \rho_q({\bf r},t). 
\label{rq}
\end{equation}
In the above equation, $N_n = N$ and $N_p = Z$ denote neutron and proton number, respectively.
As we will see, these quantities are quite useful in the following analysis (see Section III). 

Because test particles are often associated with finite width wave packets (we use triangular functions \cite{TWINGO}), 
some surface effects are automatically included in the initialization procedure and in the dynamics,
even though explicit surface terms, as those contained in the effective Skyrme interactions, are not 
considered.  This implies that, for the surface terms, one cannot simply use the coefficients associated
with the SAMi-J parametrizations.  
Indeed we observe that a good reproduction of the experimental values of the proton root mean square radius and binding energy, for the nuclei
selected in our analysis, is obtained
when taking  $C_{surf} = D_{surf} = 0$ in our parametrizations.  Thus this choice has been adopted in the following.

\begin{figure}
\begin{center}
\includegraphics*[scale=0.36]{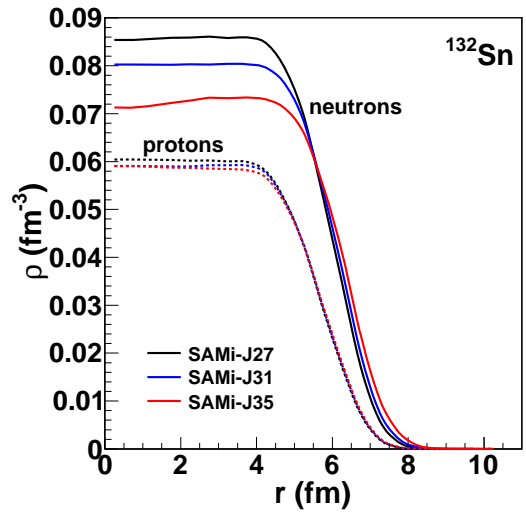}
\end{center}
\caption{(Color online) The neutron (full lines) and proton (dashed lines) density profiles of $^{132}Sn$ for the SAMi-J27, SAMi-J31 and SAMi-J35 parametrizations.} 
\label{radius}
\end{figure}

We will concentrate our analysis on three mass regions, considering the following neutron-rich nuclei:
$^{68}Ni$ (N/Z = 1.43), $^{132}Sn$ (N/Z = 1.64),  $^{208}Pb$ (N/Z = 1.54).
The corresponding values of binding energy, neutron and proton root mean square radii are reported in Tables \ref{ni68sami}, \ref{sn132sami}, \ref{pb208sami},
for the SAMi-J interactions.
In Fig.\ref{radius}, we show the neutron and proton density profiles, obtained for the system  $^{132}Sn$, with the
three SAMi-J parametrizations considered in our study. 
According to the procedure adopted here to build the ground state configuration, these values are not expected to coincide with the results of Hartree-Fock calculations, but they are actually quite close \cite{Xavi}.  
As expected, the neutron skin thickness increases
with the slope parameter $L$: this effect is indeed related to the derivative of the symmetry energy around saturation density. 
When the symmetry energy decreases significantly below $\rho_0$, as in the case of the {\it asysuperstiff} EoS
or the SAMi-J35 interaction, it is energetically convenient for the system to push the neutron excess towards
the nuclear surface.  

The same trend is observed for the $^{68}Ni$  and $^{208}Pb$ ground state configuration (see Tables \ref{ni68sami}, \ref{pb208sami}) and
also for the MI interactions \cite{coll2}.
However, it should be noticed that, in the case of the SAMi-J interactions, 
the different value of the symmetry energy at saturation induces a quite
different behavior of the neutron density also in the bulk, see Fig.\ref{radius}.

\begin{table*}[t]
\begin{center}
\begin{tabular}{|c|c|c|c|c|c|}
\hline
Interaction  & C($\rho_0$) (MeV) &  L (MeV) & Interaction  & C($\rho_0$) (MeV) &  L (MeV)  \\
\hline
asysoft & 30 & 14.8 & SAMi-J27 & 27 & 29.9  \\
\hline
asystiff & 30.5 & 79 & SAMi-J31 & 31 & 74.5 \\
\hline
asysuperstiff & 30.5 & 106 & SAMi-J35 & 35 & 115.2 \\
\hline
\end{tabular}
\caption{The symmetry energy coefficient at saturation density for 
the Skyrme interactions employed in our study and the corresponding slope $L$.}
\label{sl}
\end{center}
\end{table*}


\begin{table*}[t]
\begin{center}
\begin{tabular}{|c|c|c|c|c|}
\hline
 Interaction  & $\sqrt{\langle r^2 \rangle_n}$ (fm) &  $\sqrt{\langle r^2 \rangle_p}$ (fm) &  $\sqrt{\langle r^2 \rangle_n} -\sqrt{\langle r^2 \rangle_p}$ (fm) & BE/A (MeV) \\
\hline
SAMi-J27 & 4.043 & 3.889 & 0.154 & -9.130  \\
\hline
SAMi-J31 & 4.102 & 3.898 & 0.204 & -9.050 \\
\hline
SAMi-J35 & 4.143 & 3.900 & 0.243 & -8.971 \\
\hline
 $^{68}Ni$ Exp &   & 3.857 ( $^{64}Ni$) &  & -8.682  \\
\hline
\end{tabular}
\caption{
Neutron and proton root mean square radii, and their difference, 
and binding energy  for $^{68}Ni$, as obtained  with the SAMi-J interactions.
The experimental values, for charge radius and binding energy,  are also indicated (from \cite{jagADN1987}).}
\label{ni68sami}
\end{center}
\end{table*}


\begin{table*}[t]
\begin{center}
\begin{tabular}{|c|c|c|c|c|}
\hline
 Interaction  & $\sqrt{\langle r^2 \rangle_n}$ (fm) &  $\sqrt{\langle r^2 \rangle_p}$ (fm) &  $\sqrt{\langle r^2 \rangle_n} -\sqrt{\langle r^2 \rangle_p}$ (fm) & BE/A (MeV) \\
\hline
SAMi-J27 & 4.940  & 4.728 & 0.212 & -8.637\\
\hline
SAMi-J31 & 5.035 & 4.741 & 0.294 & -8.552  \\
\hline
SAMi-J35 & 5.150 & 4.753 & 0.397 & -8.405 \\
\hline
$^{132}Sn$ Exp &   & 4.7093 &  & -8.354  \\
\hline
\end{tabular}
\caption{The data for $^{132}Sn$, similar to Table \ref{ni68sami}.} 
\label{sn132sami}
\end{center}
\end{table*}



\begin{table*}[t]
\begin{center}
\begin{tabular}{|c|c|c|c|c|}
\hline
 Interaction  & $\sqrt{\langle r^2 \rangle_n}$ (fm) &  $\sqrt{\langle r^2 \rangle_p}$ (fm) &  $\sqrt{\langle r^2 \rangle_n} -\sqrt{\langle r^2 \rangle_p}$ (fm) & BE/A (MeV) \\
\hline
SAMi-J27 & 5.648 & 5.513 & 0.135 & -8.105  \\
\hline
SAMi-J31 & 5.735 & 5.536 & 0.198 &  -8.042 \\
\hline
SAMi-J35 & 5.813 & 5.549 & 0.264 &  -7.930\\
\hline
 $^{208}Pb$ Exp &   & 5.5012 &  & -7.867  \\
\hline
\end{tabular}
\caption{The data for  $^{208}Pb$, similar to Table \ref{ni68sami}.} 
\label{pb208sami}
\end{center}
\end{table*}

\section{Results}
\subsection{Collective Dipole Response}
We study the E1 (isoscalar and isovector) response of nuclear systems, considering initial
conditions determined by the instantaneous
excitation $\displaystyle V_{ext} =\eta_k \delta(t-t_0) \hat{D}_k$, at $t=t_0$, 
along the $z$ direction \cite{calAP1997,barPRC2012}. 
Here $\hat{D}_k$ is the isoscalar ($k=$ S) or isovector ($k=$ V) dipole operator: 
\begin{equation}
\hat{D}_S = \sum_i (r_i^2 - 5/3 \langle r^2\rangle)z_i; 
\end{equation}
\begin{equation}
\hat{D}_V = \sum_i \tau_i N/A~z_i - (1-\tau_i) Z/A~z_i,
\end{equation}
where $\tau_i =1(0)$ for protons (neutrons) and  $\langle r^2\rangle$ denotes the mean square
radius of the nucleus considered. 
According to basic quantum mechanics, 
if $|\Phi_{0} \rangle$ is the
state before perturbation, then the excited state becomes $\displaystyle |\Phi_k (t_0)\rangle 
=e^{i \eta_k \hat{D}_k} |\Phi_{0} \rangle$. The value of $\eta_k$ can be related to the initial 
expectation value of the collective dipole momentum $\hat{\Pi}_k$, which is canonically conjugated 
to the collective coordinate $\hat{D}_k$, i.e.,  $[\hat{D}_k,\hat{\Pi}_k]=i\hbar$ \cite{barRJP2012}.

For instance, in the simpler case of the 
isovector excitation, 
 $\hat{\Pi}_V$ is canonically conjugated to the collective coordinate $\hat{D}_V = (NZ/A)~\hat{X}_V$, where
 $\hat{X}_V$ defines the distance between the center of mass (CM) of protons and the CM of neutrons. 
Then one obtains:
\begin{equation}
\langle \Phi_V (t_0)|\hat{\Pi}_V|\Phi_V (t_0) \rangle  = \eta_V \frac{N Z}{A}.
\label{eta}
\end{equation}
More generally, the dipole momentum is connected to the velocity field, 
which can be extracted taking the spatial derivatives of the
perturbation $V_{ext}$ \cite{urbPRC2012}.

The strength function $S_k(E)=\sum_{n > 0}|\langle n|\hat{D}_k|0\rangle|^2\delta(E-(E_n-E_0))$, 
where $E_n$ is the excitation energy of the state $|n\rangle$ and
$E_0$ is the energy of the ground state $\displaystyle |0\rangle=|\Phi_{0} \rangle$, is obtained from the imaginary part of the Fourier transform of the time-dependent expectation value of 
the dipole moment $ \displaystyle D_k(t) = \langle \Phi_k (t) |\hat{D}_k| 
\Phi_k (t) \rangle $ as:
\begin{equation}
 S_k(E) =\frac{Im(D_k(\omega))}{\pi \eta_k }~~,
\label{stre}
\end{equation}
where $\displaystyle D_k(\omega) =\int_{t_0}^{t_{max}} D_k(t) e^{i\omega t} dt$. Since ($\pi \eta_k$) is a constant, we plot $Im(D_k(\omega))$ instead of $S_k(E)$ in the following.

Dipole oscillations and response functions
can be investigated, within our semi-classical treatment, considering  a gentle perturbation of the
ground state configuration of the nucleus under consideration and then looking at its dynamical evolution, as given by 
Eq.(\ref{vlasov}).    We follow the dynamics of the system until $t_{max}=1800$~fm/c, thus being able to extract time oscillations
of the dipole moments.  
A filtering procedure, as described in \cite{reiPRE2006}, was applied in order to eliminate the artifacts resulting from a finite time domain analysis of the signal. Thus a smooth cut-off function was introduced such
that $D_k(t) \rightarrow D_k(t)\cos^{2}(\frac{\pi t}{2 t_{max}}) $. 

\begin{figure}
\begin{center}
\includegraphics*[scale=0.36]{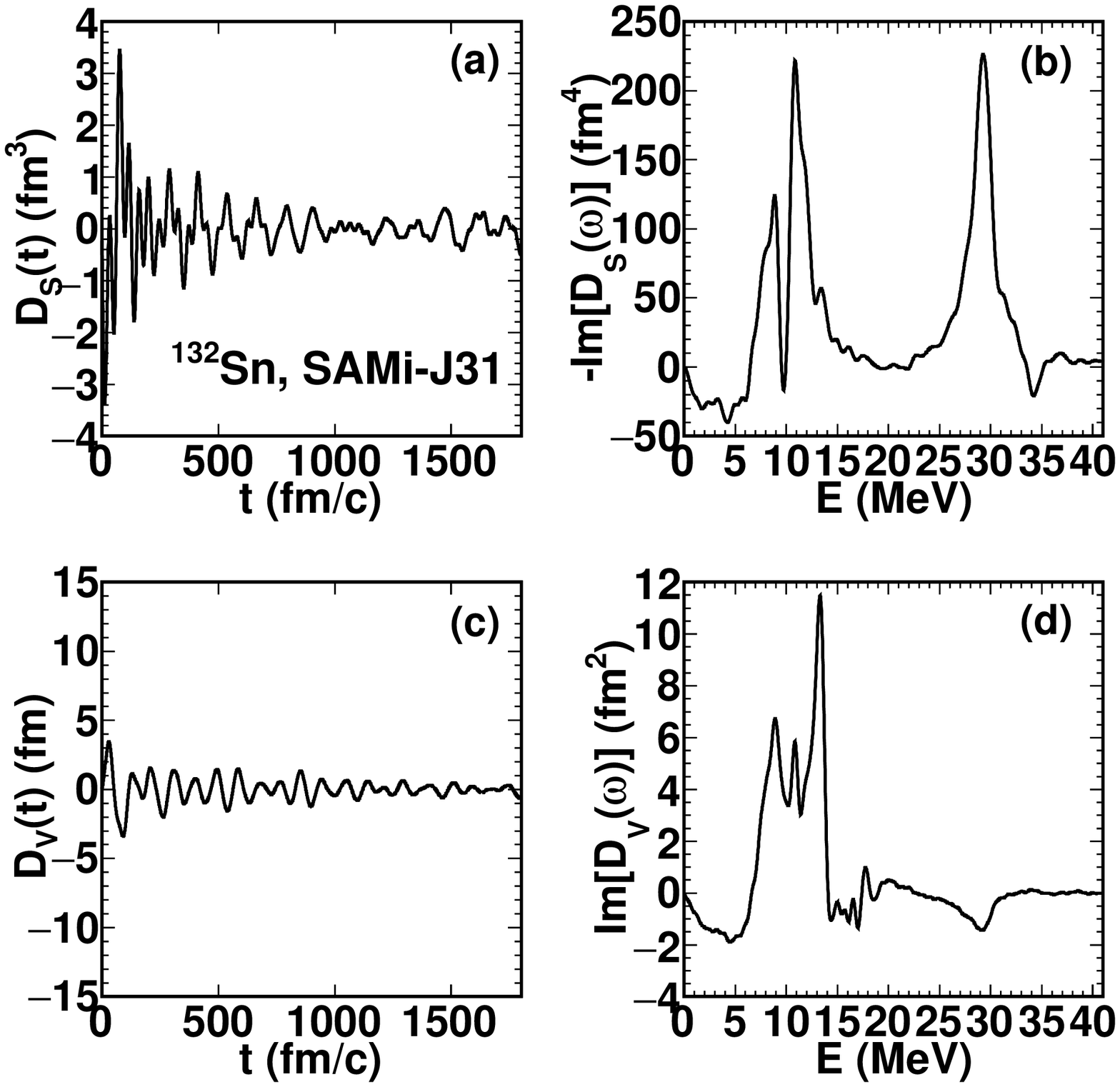}
\includegraphics*[scale=0.36]{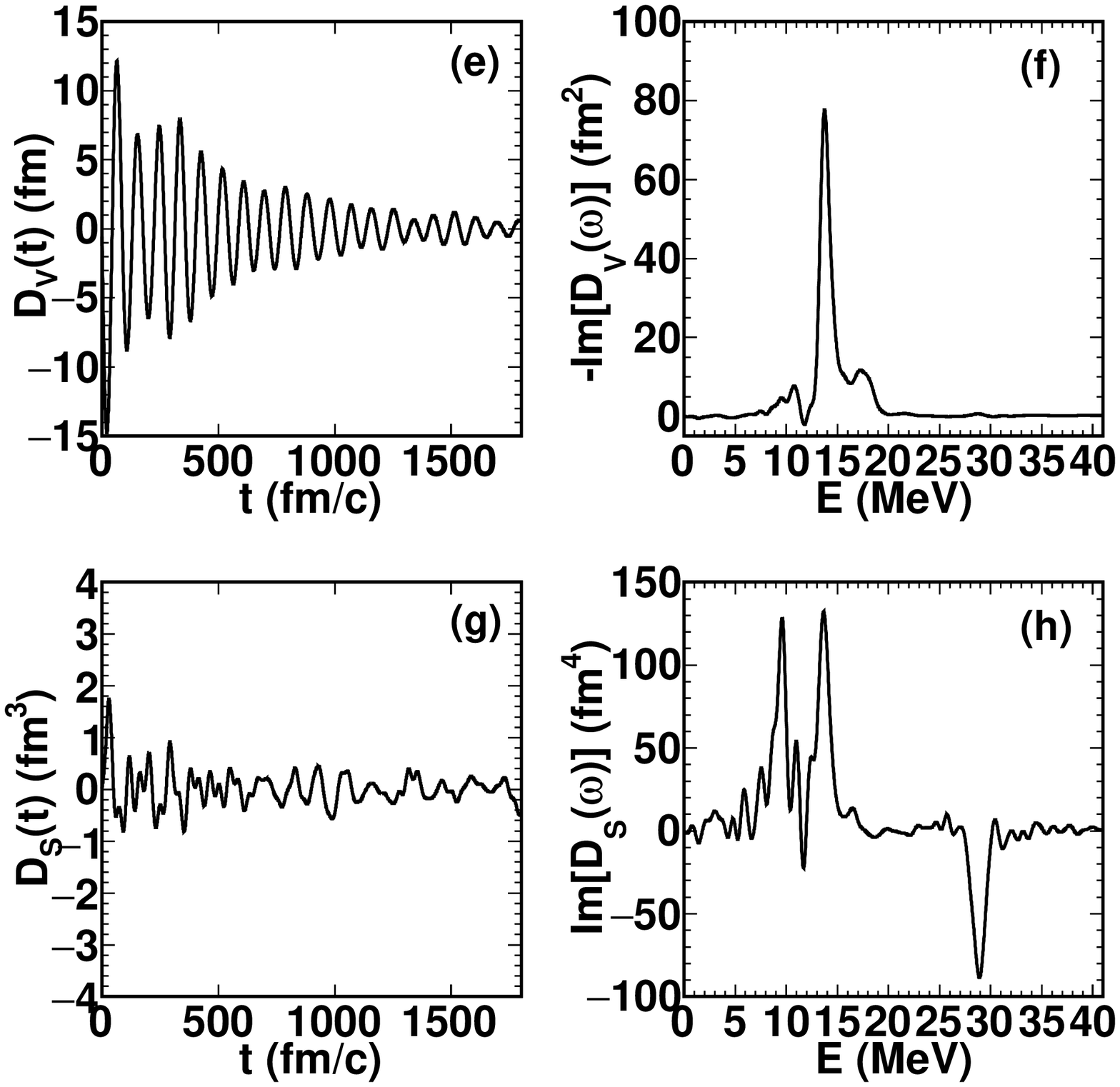}
\end{center}
\caption{The dipole oscillations (left panels) and corresponding strength (right panels) for $^{132}Sn$ and the SAMi-J31 interaction. Panels from (a) to (d) represent the results obtained with the initial IS perturbation and panels from (e) to (h) show the results obtained with the initial IV perturbation.}
\label{isivsn132sami31}
\end{figure}

As it is well known, in symmetric nuclear matter isoscalar and isovector modes are fully decoupled.  However, in
neutron-rich systems, neutrons and protons may oscillate with different amplitudes, thus inducing a coupling of
isoscalar and isovector excitations. 
One of the main goals of our analysis is to get a deeper insight into this effect.
Indeed it appears that considering an initial isovector perturbation of the system, one also gets an isoscalar response, and vice versa. 
This is illustrated  in Fig.\ref{isivsn132sami31}, where we represent dipole oscillations (left panels) and 
corresponding strength, as a function of the excitation energy $E = \hbar \omega$ (right panels) for the system $^{132}Sn$ and the SAMi-J31 interaction,  
obtained by considering an initial IS perturbation with  
$\eta_S = 0.5$~MeV~fm$^{-2}$ (panels from (a) to (d)) or an initial IV perturbation with 
$\eta_V = 25$~MeV (panels from (e) to (h)).

One can observe that, when introducing an IS perturbation
at the initial time $t_0$ (Fig.\ref{isivsn132sami31}, panels (a)-(b)), 
also isovector-like modes are excited, as it is evidenced from the analysis
of the corresponding isovector dipole oscillations and associated strength (panels (c)-(d)). 
Similarly, an initial IV perturbation (panels (e)-(f))
also generates an isoscalar response (panels (g)-(h)).

In the isovector response (panel (f)) one can easily recognize the main 
IV GDR peak, with $E_{GDR}\approx 14$ MeV.  Some strength is also evidenced at lower
energy (mostly in the range between $E_1 = 9$ MeV and $E_2 = 11$ MeV), which could be associated with the PDR.  
These low-energy modes contribute significantly to the
corresponding 
isoscalar projection (panel (h)), now acquiring a larger strength, 
comparable to that associated with the robust GDR mode, thus manifesting their
isoscalar-like nature.  
A (negative) peak is seen at higher energy (around 29 MeV), which
corresponds to the giant isoscalar-like dipole mode (IS GDR) \cite{mazPRC2012}
which is also excited, owing to its mixed character, by the initial perturbation. 

When agitating the system  
with an initial isoscalar excitation,
essentially the same oscillation modes emerge, with a larger strength 
for the isoscalar-like ones in this case.
Indeed, in the isoscalar response (panel (b))
two main peaks, whose positions are quite close to the $E_1$ and $E_2$ energies evidenced
in panel (h), are observed in the low energy region, 
together with some (smaller) strength located around the IV GDR region ($E_{GDR} \approx 14$~MeV). A quite large contribution appears also in the high energy region of the spectrum ($E \approx 29$~MeV). 
Projecting onto the isovector direction (panel (d)) the strength of the IV GDR mode
is enhanced, as expected according to its isovector-like nature, becoming comparable 
to that of the low-energy isoscalar-like modes excited by the initial perturbation.  
On the other hand,  
the high energy mode exhibits a quite small (negative) strength, pointing again to its isoscalar-like character. 

To summarize, we observe that the same energy modes, which are actually the normal
modes of the system and are of mixed nature, appear at the same time in
the isoscalar and isovector responses of the system, but with a different weight, 
depending on their intrinsic structure and on the initial perturbation type. 
In particular, the low-energy modes, lying below the GDR peak, have 
predominant isoscalar nature, but they may also contribute to the isovector 
response, in the PDR region. 

We move now
to investigate how the response of the system depends
on the effective interaction adopted, in the three mass regions considered in 
this work. 
Hereafter we will only examine the isoscalar (isovector)  response connected 
to an initial isoscalar (isovector) perturbation.  
\begin{figure}
\begin{center}
\includegraphics*[scale=0.36]{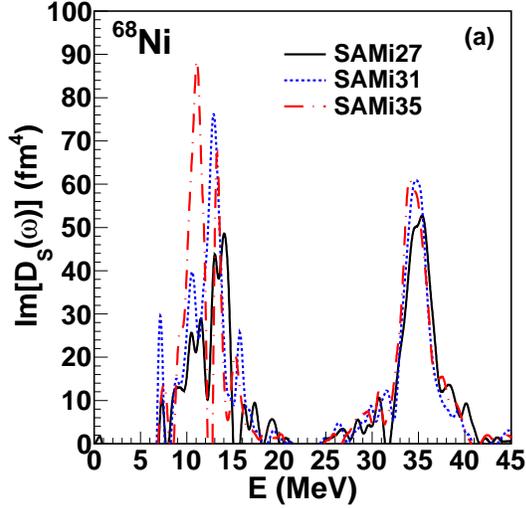}
\includegraphics*[scale=0.36]{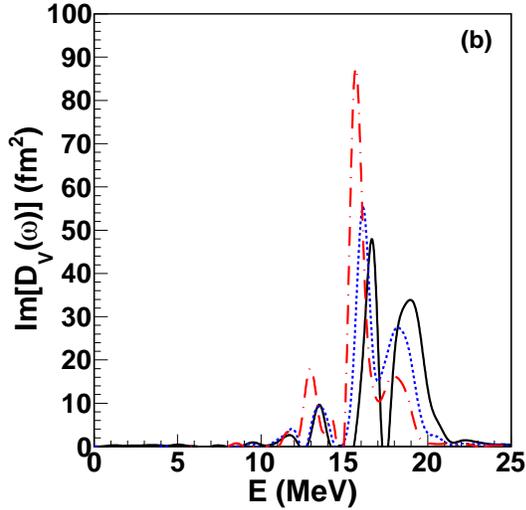}
\end{center}
\caption{(Color online) The strength functions versus excitation energy for $^{68}Ni$ with SAMi-J27, SAMi-J31 and SAMi-J35 interactions. Panel(a) is for the initial IS perturbation and panel (b) is for the initial IV perturbation.}  
\label{ni68isiv}
\end{figure}

\begin{figure}
\begin{center}
\includegraphics*[scale=0.36]{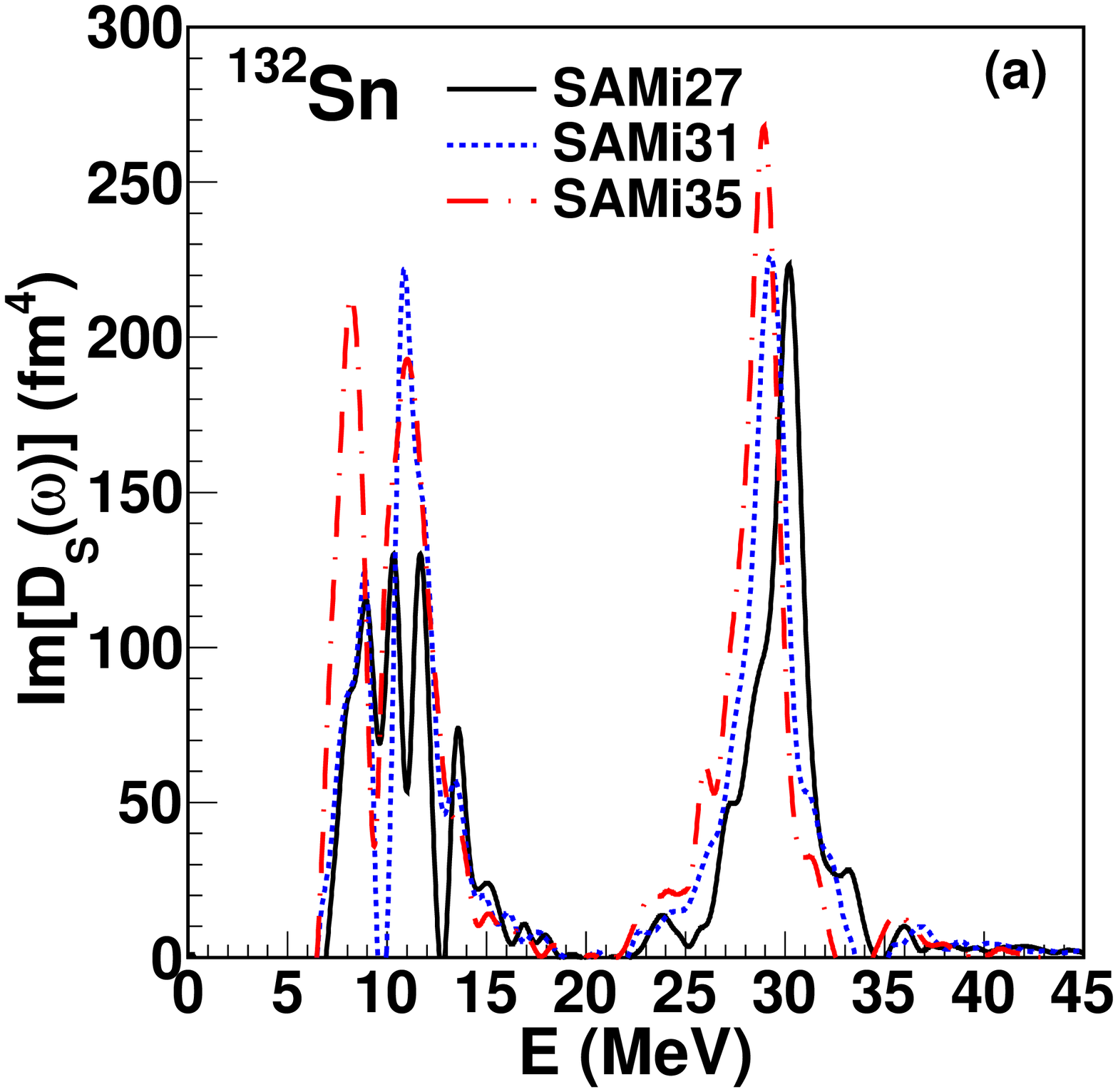}
\includegraphics*[scale=0.36]{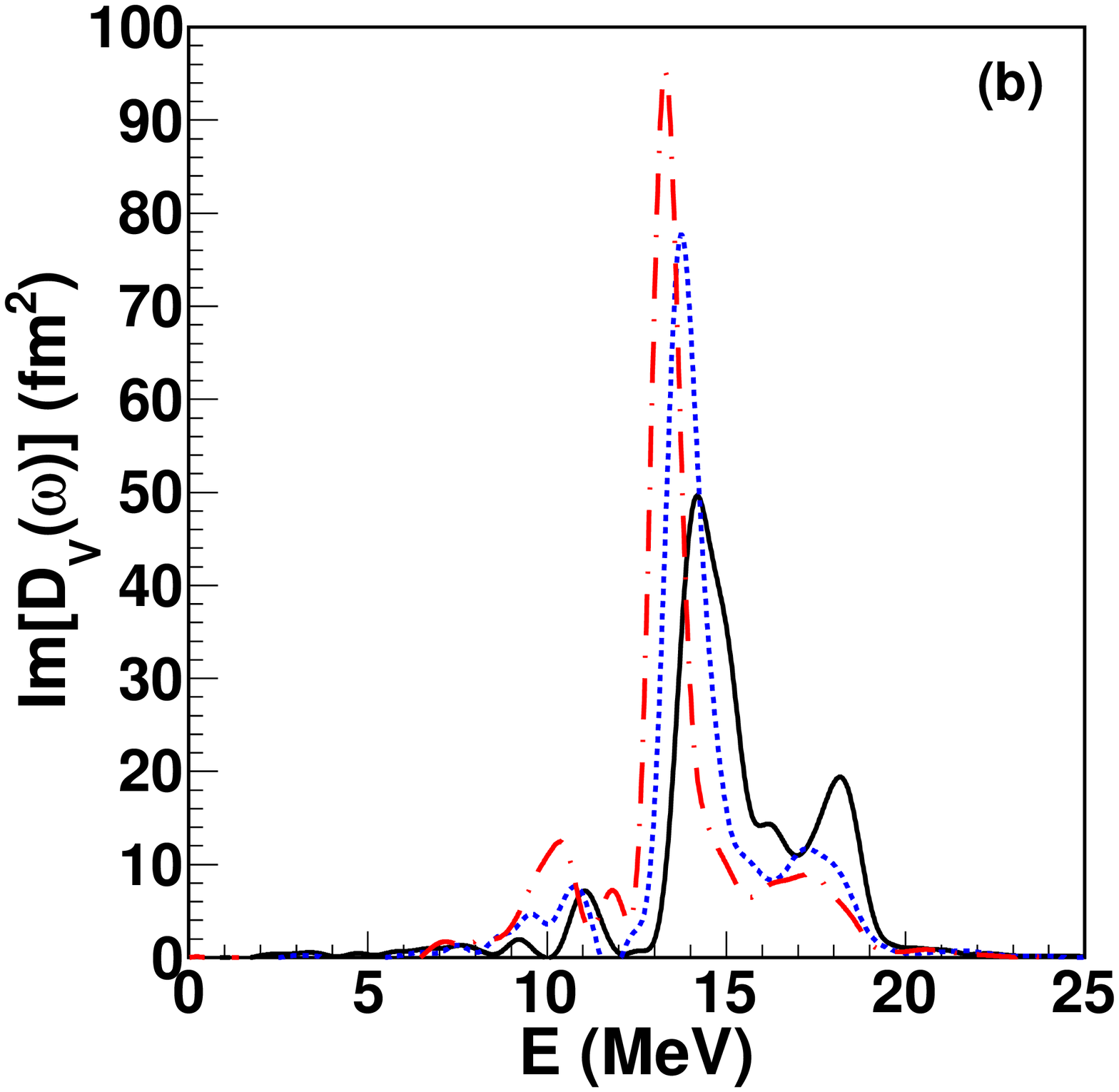}
\end{center}
\caption{(Color online) Similar to Fig.\ref{ni68isiv} but for $^{132}Sn$.}
\label{sn132isiv}
\end{figure}

\begin{figure}
\begin{center}
\includegraphics*[scale=0.36]{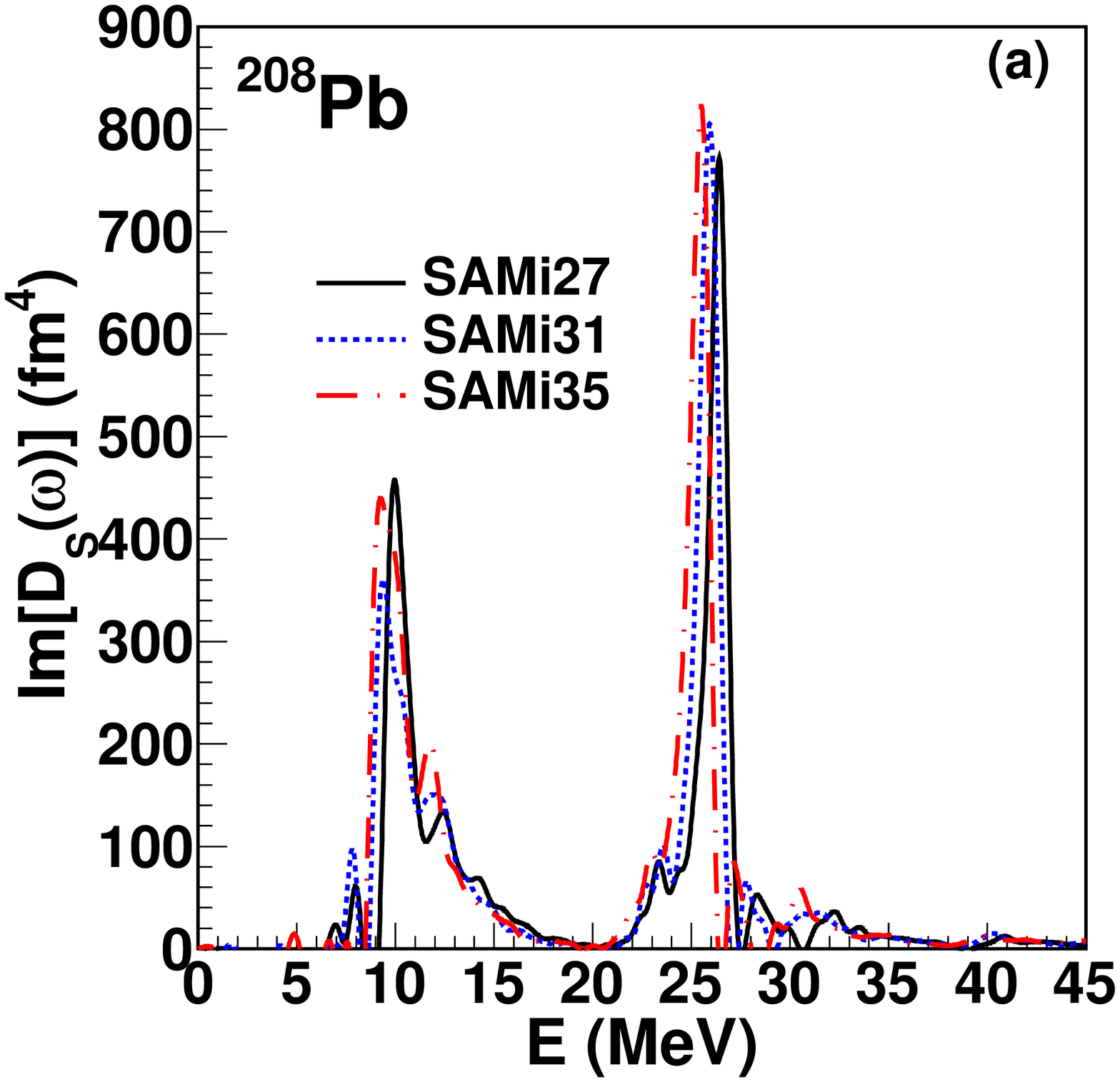}
\includegraphics*[scale=0.36]{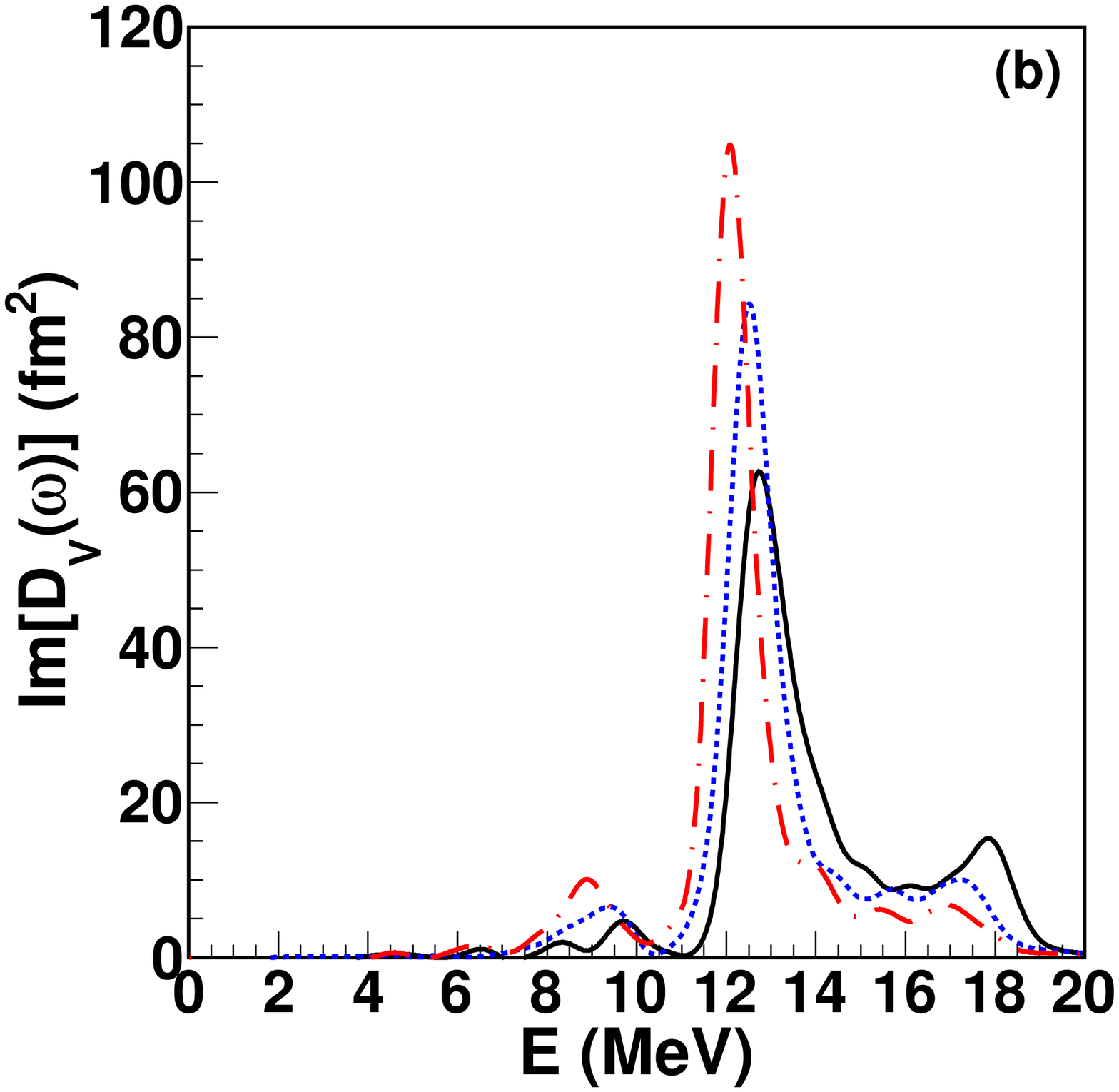}
\end{center}
\caption{(Color online) Similar to Fig.\ref{ni68isiv} but for $^{208}Pb$.}
\label{pb208isiv}
\end{figure}

\begin{figure}
\begin{center}
\includegraphics*[scale=0.36]{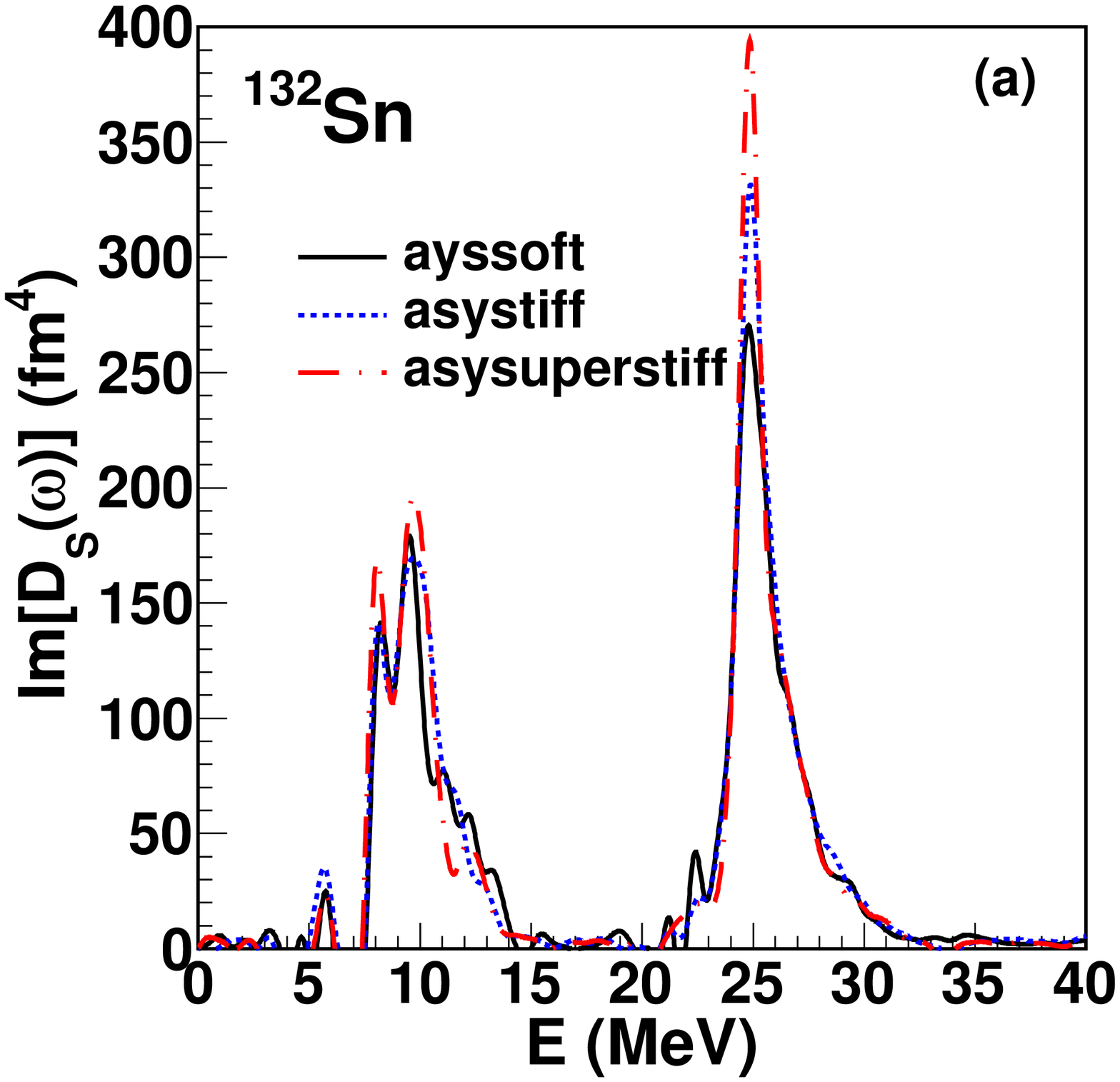}
\includegraphics*[scale=0.36]{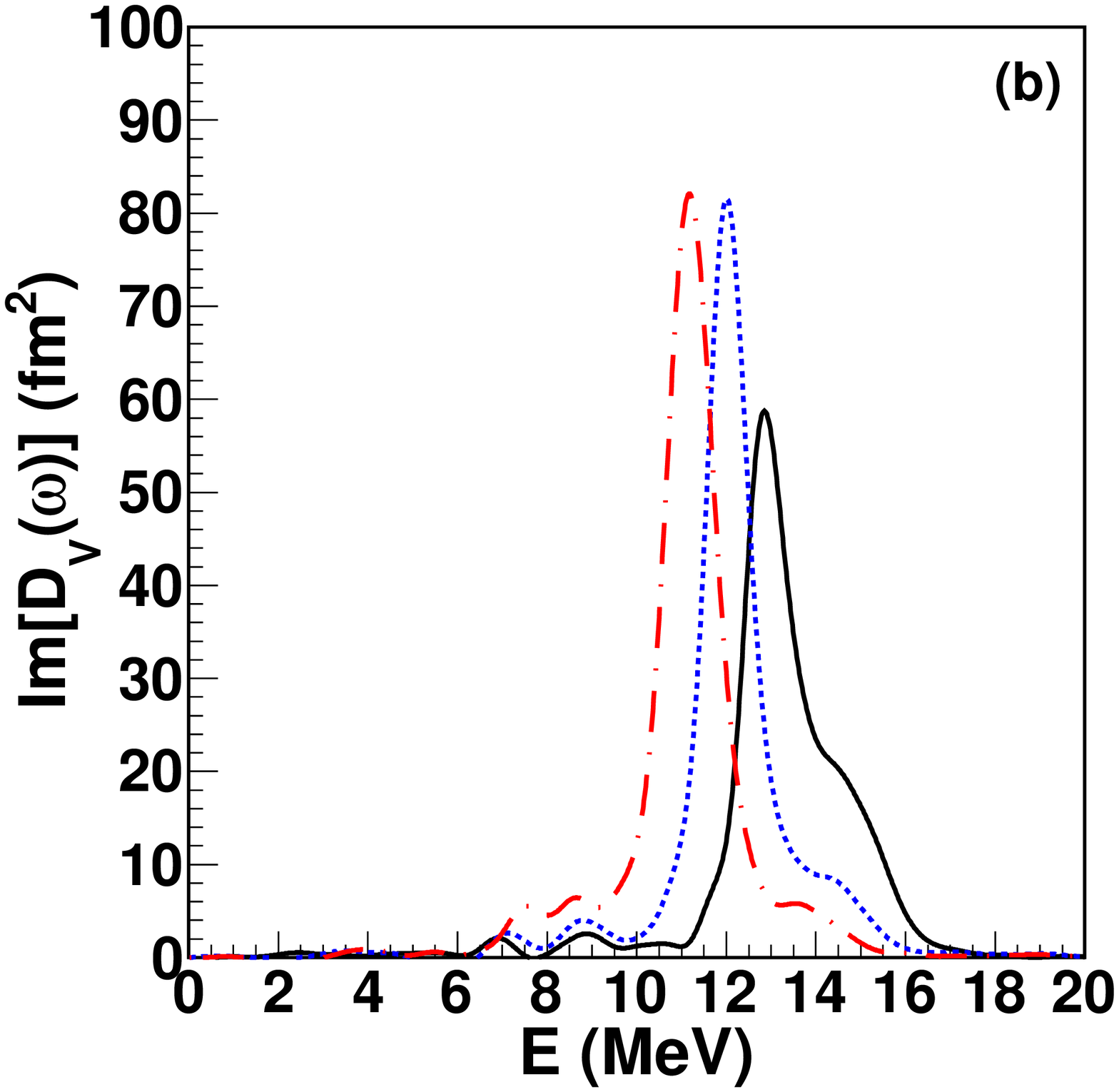}
\end{center}
\caption{(Color online) Similar to Fig.\ref{sn132isiv}, but for the MI interactions.}
\label{sn132isivorig}
\end{figure}

In Fig.\ref{ni68isiv} (a)
we show, for $^{68}Ni$, the strength function corresponding to the IS
dipole response as a function of the excitation energy $E$. In Fig.
\ref{ni68isiv}(b),
the same quantity is shown but for the IV dipole response.
In the case of $^{132}Sn$ and
$^{208}Pb$, the strength functions for the
dipole response are depicted in Figs. \ref{sn132isiv}(a) and
\ref{pb208isiv}(a) (IS) and in
Figs. \ref{sn132isiv}(b) and \ref{pb208isiv}(b) (IV), respectively.
In all
panels, the predictions of the three selected SAMi-J interactions are
shown. 


For $^{68}Ni$ and $^{132}Sn$, 
the isoscalar strength (panels (a)) appears quite fragmented in the low-energy domain. However one
can recognize two main regions of important contribution for all the interactions considered (see in particular the SAMi-J35 results)
and identify a smaller peak centered at the energy of the IV GDR (originating from its mixed nature in 
neutron-rich systems, as stressed above). 
It is worth noting that the observation of two main low-energy peaks
in the isoscalar response is in agreement with the semi-classical studies of
Ref.\cite{urbPRC2012}, 
where isoscalar toroidal excitations are investigated.  
In particular, in \cite{urbPRC2012} it is shown that the lowest energy mode is associated mostly
with surface oscillations and, in the case of neutron-rich systems, 
is responsible for the low-lying strength observed in the isovector
response (in the PDR region).  
On the other hand, RPA calculations \cite{mazPRC2012} exhibit a more isolated
peak in the low-energy region of the IS strength function $S_S(E)$, but 
some contributions appear also at higher energy, in a region around
the domain of the IV GDR (for instance, around 14 MeV in the  $^{132}Sn$ case).

For the largest system considered,  $^{208}Pb$, our calculations show 
just one main peak, of significant strength, in the low-energy region. 

The discrepancy with respect to RPA calculations may be probably
attributed  to the lack of intrinsic gradient terms of quantal nature 
in our approach and to
the numerical treatment of surface effects \cite{Ayik}.
It appears more
critical in smaller systems, where the relative importance of surface to
volume effects increases. 

In any case, the low-energy peaks of the IS response appear connected  to the low-lying strength
observed in the IV response (panels (b)), in the PDR region. 
Since 
the different peaks are quite close to each other, only one main peak,
resulting from two interfering contributions, 
may appear
in the PDR 
region of the IV response. 

Let us concentrate now on the details of the isovector response. 
In the $^{208}Pb$ case, 
the centroid energies of the PDR as well as the energy peak
of the isovector GDR
predicted by the employed interactions (E=8-10 MeV and E=12-13 MeV, respectively) are close to
the experimental data (E=7.37 MeV within a
window of 6 - 8 MeV \cite{pb1} and
E=13.43 MeV with a total width of 2.42 MeV \cite{ber1975}, respectively).
The predictions
of the three SAMi-J interactions for the PDR, for
$^{132}Sn$ (E= 9.0-11.0 MeV) and 
for $^{68}Ni$ (E= 11.5 - 13.5 MeV), are also close, but still a little higher than
the measured data (E=
9.1 - 10.5 MeV for
$^{132}Sn$ \cite{adrPRL2005}
and E=11 MeV with an energy width estimated to be less
than 1 MeV for
 $^{68}Ni$ \cite{wiePRL2009,ni3}). 
The overestimation of the PDR energy in our calculations may still be
connected to 
the semi-classical treatment of surface effects, as already stressed above.
Indeed the PDR region is essentially populated by low-lying isoscalar-like
oscillations, whose energy is significantly affected by surface effects.  
The results
can be probably improved by a fine tuning of the
coefficients $C_{surf}$ and  $D_{surf}$ in the Skyrme parametrizations. 

Qualitatively, in the three 
nuclei it appears that the larger the value of $L$, 
the higher the
different peaks arising in the low-energy region 
of the IV dipole response (see Figs. \ref{ni68isiv} - \ref{pb208isiv}, panels (b)).
Moreover, as it clearly appears from panels (a), the strength
of the lowest energy mode in the IS response increases (except for $^{208}Pb$)  when increasing the slope $L$
of the parametrization considered, also reflecting into a 
larger isovector strength in the PDR region (panels (b)). 
We note that, on the basis of nuclear matter calculations \cite{barPR2005},
we expect a larger degree of mixing between isoscalar and isovector modes, 
in neutron-rich systems, for symmetry energy parametrizations with 
larger slope $L$.  Moreover, in this case, one also obtains a more 
extended neutron skin (see Fig.\ref{radius}), thus surface and isospin effects 
are both enhanced.

Finally, we observe for all nuclei that the IV projection of the PDR is
an order of magnitude smaller than the IV GDR, but its
isoscalar counterpart is of the same order of magnitude as
the corresponding IS GDR \cite{mazPRC2012}. We conclude that the PDR is mostly an isoscalar low-energy mode, 
involving also nucleons which belong to the nuclear surface. Owing
to the charge asymmetry of the systems considered, this mode also manifests
an isovector character, especially in the case of the stiffer interactions, 
which predict a larger asymmetry in the surface (neutron skin),
see the density profile in Fig.\ref{radius}.


Actually the IV response also exhibits other interesting features, 
which can be better discussed by comparing with 
the results obtained with the MI Skyrme
interactions, 
displayed in Fig.\ref{sn132isivorig} for  $^{132}Sn$.
Concerning the main IV GDR, our calculations indicate that its excitation
energy is mainly affected by the value of the symmetry energy at the density
$\rho_c = 0.65~\rho_0$, where the three SAMi-J interactions 
cross each other (see Fig.1), which can be taken as 
the average density of medium-heavy nuclei.  
Indeed, the centroid of the IV GDR peak does not evolve much with the
parametrization considered, in the SAMi-J case.  
The largest shift is observed for the smallest system, $^{68}Ni$, indicating that the
GDR centroid is actually sensitive to the value of the symmetry energy at a density below $\rho_c$ 
in this case.   
On the other hand, 
for the MI parametrizations, which cross at normal density (see Fig.1),
thus having a smaller value of the symmetry energy below $\rho_0$ in the 
stiffer case, 
the energy centroid  is clearly more sensitive to the parametrization
employed, see Fig.\ref{sn132isivorig}, being smaller in the $asysuperstiff$ case.   
We also stress that the GDR energy appears always underestimated by
the MI interactions, whereas it is close to the experimental observation
when the SAMi-J interactions are considered.
In particular, 
the SAMi-J31 and the asystiff parametrizations 
are characterized by a quite similar behavior of the
symmetry energy (compare the two panels of Fig.1), nevertheless
the results of the dipole response are
different in the two cases.  This highlights the role of momentum dependent effects
in shaping the features of the nuclear response.   
One can also note that the energy location of the PDR strength is much less
sensitive to the isovector channel of the interaction, especially in the MI case.

In the isovector dipole response obtained with the SAMi-J, we also observe a quite pronounced peak at higher energy, with respect to the GDR, whose strength decreases
with the stiffness of the interaction, in agreement with 
RPA calculations \cite{mazPRC2012}. 
This peak is less pronounced in the MI case.  

\subsection{Transition densities}
In addition to the investigation of the dipole response presented above, 
the analysis of
the transition densities associated with the different excitation modes of the system is very instructive since 
it delivers important information about the spatial structure related to the dynamics of every excitation. 

To undertake this analysis, we need to evaluate the local spatial density as a function of time.  
In order to reduce numerical fluctuations, 
we take into account the 
cylindrical symmetry of the initial perturbation and, averaging over the azimuthal 
$\phi$ angle, we extract the density $\rho_q(r,\cos\theta,t)$ and the corresponding
fluctuation  $\delta\rho_q(r,\cos\theta,t) 
= \rho_q(r,\cos\theta,t) - \rho_q(r, t_0)$, where $\cos\theta = z/r$ and
$\rho_q(r, t_0)$ denotes the ground state density profile which only depends
on $r$.

As suggested in Ref.\cite{urbPRC2012}, 
assuming that the amplitude of the oscillation is weak (linear
response regime), the spherical symmetry of the ground
state and the dipole form of the excitation operator imply that the
transition density can be written, at each time, as:
$\delta \rho_q(r,\cos\theta,t)=\delta\rho_q(r,t)\cos\theta$.
Then one can finally extract the
transition density just as a function of the radial distance $r$, by averaging,  over $\cos\theta$,
the quantity  $\delta \rho_q(r,t) = \delta\rho_q(r,\cos\theta,t) /
\cos\theta$.


It is clear that 
the delta function perturbation, $V_{ext}$, at $t = t_0$, agitates simultaneously all
modes which can be excited by the operator $\hat{D}_k$. 
Thus the corresponding density oscillations observed along the dynamical 
evolution will appear as the result of the combination of the different excitation modes. 
In order to pin down the contribution of a given mode to the density 
oscillations, one can consider the energy $E$ associated, for instance, with
a peak in the strength function and compute the transition
density as the Fourier transform of
$\delta{\rho}_q(r,t)$: 
\begin{equation} \delta{\rho}_q(r,E) \propto
  \int_{t_0}^\infty dt  \delta{\rho}_q(r,t) 
    \sin\frac{E t}{\hbar} . 
\end{equation}
In practice, since the simulation runs
only to $t_\mathrm{max} = 1800$ fm$/c$, the sine 
function is multiplied by a damping factor, 
as in the strength function $S_k(E)$. 

\begin{figure}
\begin{center}
\includegraphics*[scale=0.36]{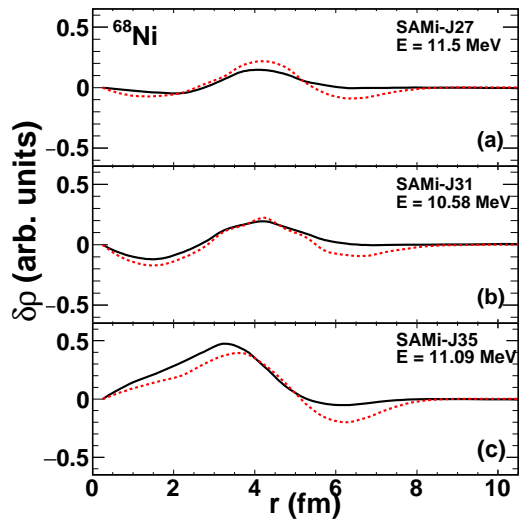}
\end{center}
\caption{(Color online) The transition densities versus $r$ in the low energy excitation region, for the IS initial perturbation, for $^{68}Ni$ with SAMi-J27, SAMi-J31 and SAMi-J35 interactions.
 Full lines are for protons, dashed lines for neutrons.
The energy of the excitation mode considered is indicated in each panel.}
\label{Trhoni68samiIS}
\end{figure}

\begin{figure}
\begin{center}
\includegraphics*[scale=0.36]{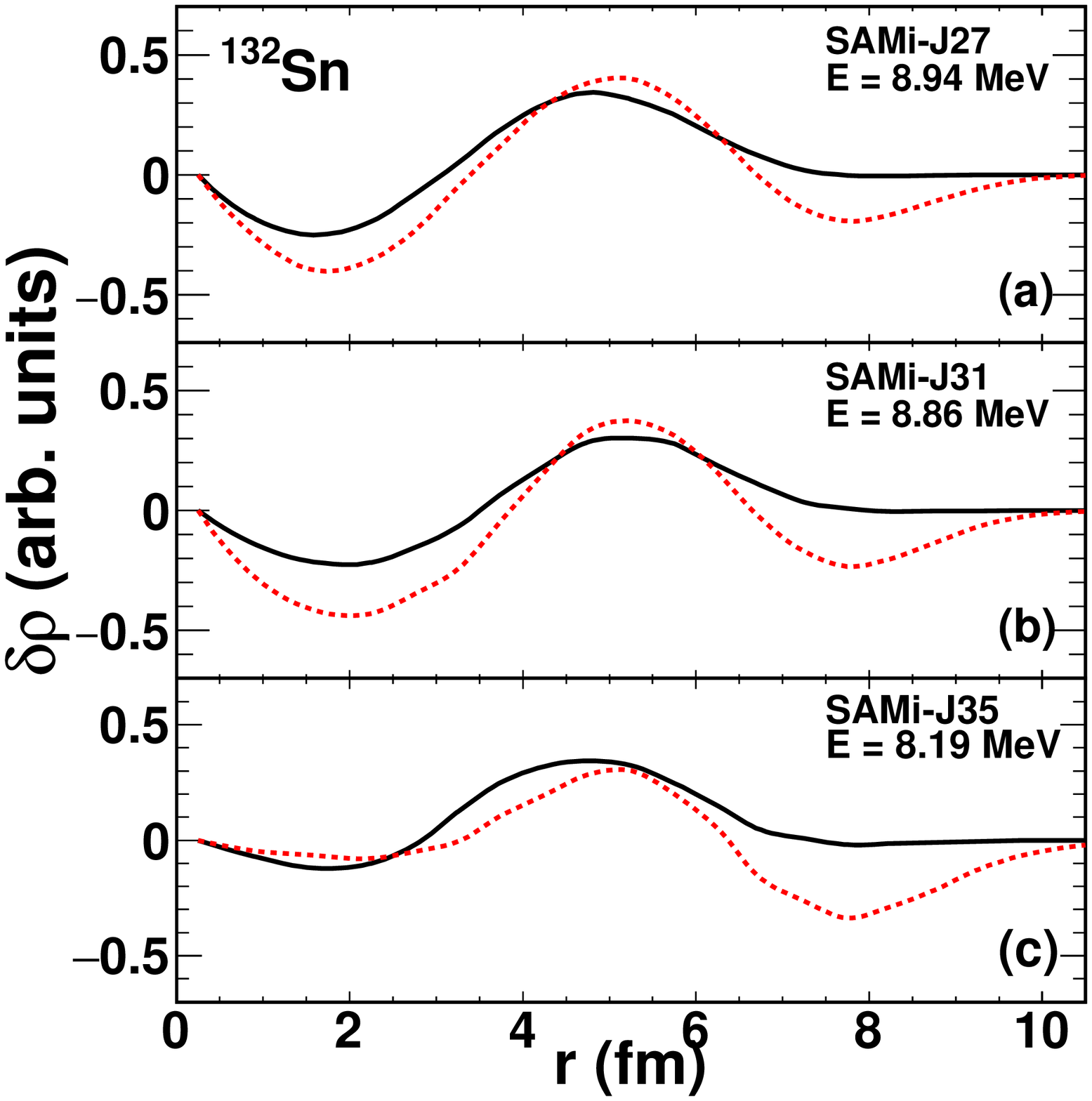}
\end{center}
\caption{(Color online) Similar to Fig.\ref{Trhoni68samiIS} but for $^{132}Sn$.}
\label{Trhosn132samiIS}
\end{figure}

\begin{figure}
\begin{center}
\includegraphics*[scale=0.36]{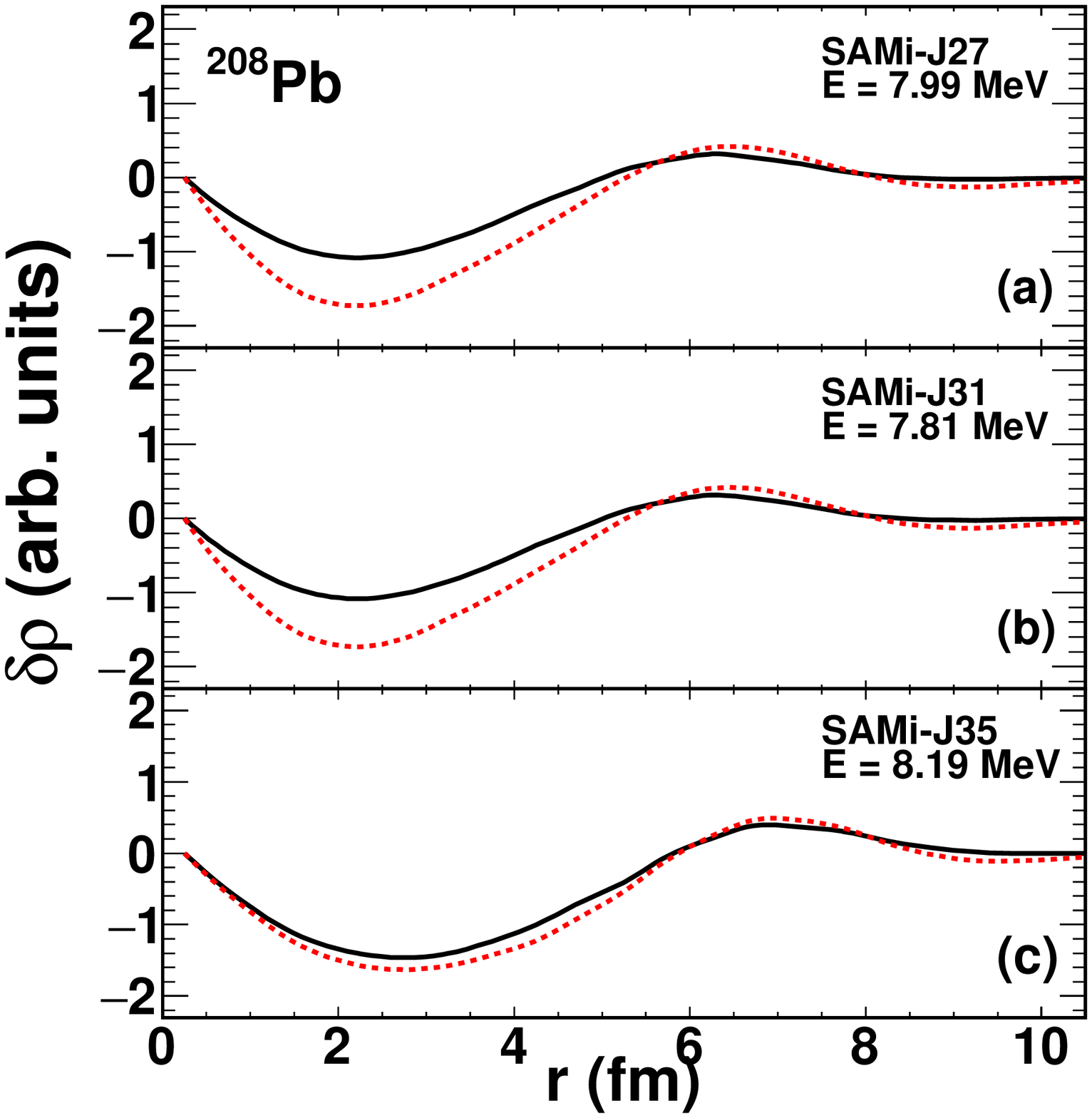}
\end{center}
\caption{(Color online) Similar to Fig.\ref{Trhoni68samiIS} but for $^{208}Pb$.}
\label{Trhopb208samiIS}
\end{figure}

\begin{figure}
\begin{center}
\includegraphics*[scale=0.36]{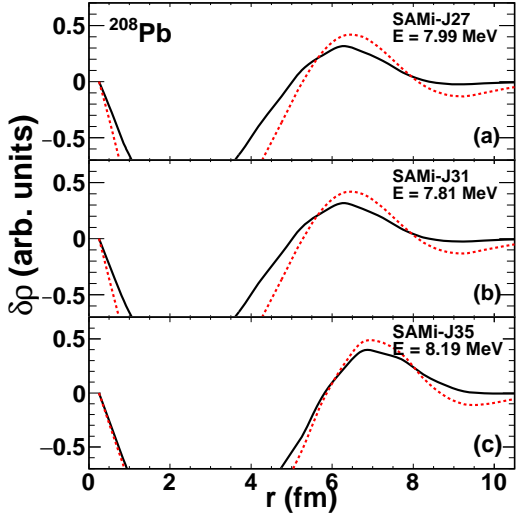}
\end{center}
\caption{(Color online) Same as Fig.\ref{Trhopb208samiIS} but with 
a reduced scale on the vertical axis.}
\label{Trhopb208samiISzoomin}
\end{figure}

\begin{figure}
\begin{center}
\includegraphics*[scale=0.36]{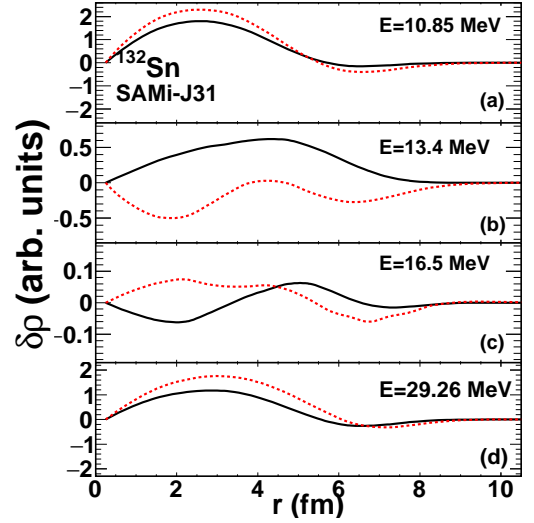}
\end{center}
\caption{(Color online) The transition densities versus $r$ with different excitation energies with IS initial perturbation for $^{132}Sn$ with SAMi-J31 interaction. Full lines are for protons, dashed lines for neutrons.}
\label{Trhosn132sami31IS}
\end{figure}

\begin{figure}
\begin{center}
\includegraphics*[scale=0.36]{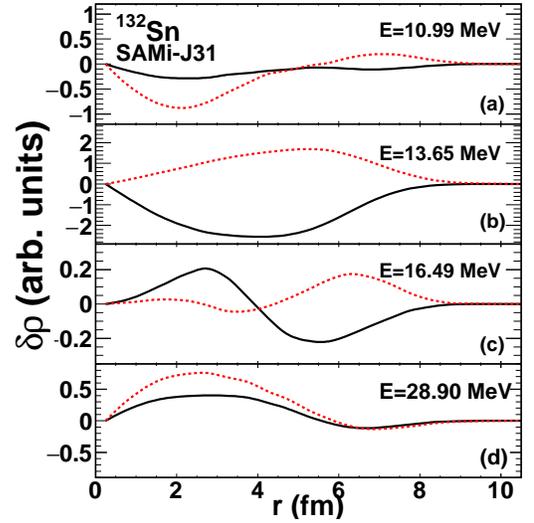}
\end{center}
\caption{( Color online) Similar to Fig.\ref{Trhosn132sami31IS} but with IV initial perturbation.}
\label{Trhosn132sami31IV}
\end{figure}


To further analyze the isoscalar or isovector character of each excitation
mode, we calculate neutron and proton transition densities. It is well known that in symmetric matter, neutrons and protons oscillate with exactly equal
(isoscalar) or opposite (isovector) amplitudes.  In neutron-rich systems,
the picture is more complex, however one can still identify isoscalar-like
modes, when the two nuclear species oscillate in phase, and isovector-like
modes, with neutrons and protons oscillating out of phase.
Apart from this information, connected to the mixed character of each mode, 
the overall spatial structure of the transition densities tells us which
part of the system (internal part, surface) is more involved in the oscillation.   

In Figs.\ref{Trhoni68samiIS}-\ref{Trhopb208samiIS}, we represent the transition density associated with the low energy
peaks 
observed in the isoscalar response, for the three systems considered and
the three SAMi-J parametrizations adopted. As discussed above, owing to its mixed character,
this mode also contributes to the isovector response, in the PDR region.
The corresponding transition density could also be extracted from the isovector response, however, since the IV strength is quite small, 
numerical fluctuations would spoil the signal \cite{romanian}. 

We observe that neutrons and protons oscillate in phase, but with different amplitudes,
with neutrons having generally larger amplitude than protons.  
The nuclear surface is significantly involved in these oscillations. Moreover, 
when considering interactions with increasing slope $L$ (from SAMi-J27 to SAMi-J35),
one can see that neutron oscillations become  larger, with respect  
to proton oscillations, especially in the surface region, whereas the opposite
seems to hold for the interior of the system.  
This can be explained by the fact that, for increasing $L$, the system asymmetry
is more pushed towards the surface, corresponding to the development of the
neutron skin, whereas the internal part  of the system becomes more symmetric. 
As one can see from Fig.\ref{Trhopb208samiISzoomin}, where the surface region of the transition density is better evidenced, surface effects are less pronounced
in the $^{208}Pb$ case. 
However, a significant contribution to the dipole strength may also come from
the intermediate spatial region, where the transition densities are positive.
Indeed, according to the definition of the IV dipole moment, Eq.(7), 
the dipole strength increases when $\delta\rho_n/\delta\rho_p > N/Z$, for negative transition densities, or when  $\delta\rho_n/\delta\rho_p < N/Z$, for positive transition densities.   Both conditions are better satisfied, in the surface and in the 
intermediate region respectively, with increasing $L$.  
This determines an overall increase of the mixed character of the mode, mainly
determined by the surface behavior, but also by the internal part of the system,  
leading to a larger strength observed in the isovector response, 
see Figs.\ref{ni68isiv}-\ref{pb208isiv}.



We also extend our analysis to the other modes giving a relevant contribution
to the isoscalar and the isovector responses. 
This is illustrated in Figs.\ref{Trhosn132sami31IS}-\ref{Trhosn132sami31IV}, for the system $^{132}Sn$, in the case of
the SAMi-J31 interaction, for IS and IV excitations, respectively. 

As it is observed from the analysis of the isoscalar response (Fig.5), 
there exists a second mode, around $E_2 = 11$ MeV, 
which gives an important contribution in the low-energy region. This excitation also
contributes to the isovector response, as already stressed in Section III.A.  
Looking at the associated transition density, generated by an IS
perturbation of the system (Fig.\ref{Trhosn132sami31IS}, panel (a)), it appears that neutrons and 
protons essentially move in phase, but still with different amplitudes. 
Thus the oscillation has a mixed character and this is why it presents
some strength in the isovector response.
Now the interior of
the system is more involved in the oscillation, though the surface is still affected.
It is worth noting that,
also for this transition density, the asymmetry increases, with $L$,  at the surface and diminishes in the internal part. 
 
We observe that, when this energy region is excited from the IV
operator (Fig.\ref{Trhosn132sami31IV}, panel (a)), though the structure of the mode keeps similar,  
the difference between neutrons and protons becomes more
pronounced. 
This effect
could be due to the influence of the strong isovector oscillations associated
with the IV GDR region, whose contribution may extend to the considered energy.
Indeed, it should be noted that even if the energy $E$ corresponds to a peak
in $S_k(E)$, the transition densities obtained with
the method employed here may still contain contributions from other modes 
if those
have a width which makes their spectrum extend to the energy $E$
\cite{urbPRC2012}.  

The highest energy isoscalar mode, that should be associated with  the isoscalar giant dipole 
compression mode, corresponds to transition densities which affect significantly
the interior of the system (Figs.\ref{Trhosn132sami31IS}-\ref{Trhosn132sami31IV}, panel(d)) and its features do not 
depend much on
the type of initial perturbation. Moreover it appears of quite robust isoscalar nature,
with a small isoscalar/isovector mixing, especially at the surface. 

It is also interesting to look at the modes which are isovector-like. In this case neutrons and protons oscillate mostly out of phase,
with protons having larger amplitude.  The transition densities extracted
from the isoscalar or from the isovector responses exhibit similar 
features, compare panels (b) and (c) in Figs.\ref{Trhosn132sami31IS}-\ref{Trhosn132sami31IV}.    
It appears that the main IV GDR mode (panels (b)) corresponds essentially to
 one oscillation, with a 
maximum close to the nuclear surface.  This result is compatible with GT picture 
of neutron and proton spheres oscillating against each other. 
On the other hand, the higher energy peak, $E \approx 16.5$ MeV (panels (c)), corresponds to 
a kind of double oscillation, which is typical of SJ
modes, i.e. volume oscillations, involving also the internal part of the system.

\section{Conclusions}







  

In this work we have addressed some of the open questions 
concerning the nature of the low-lying IV dipole strength experimentally observed in neutron-rich nuclei \cite{savPPNP2013}.
By performing a systematic investigation over three mass regions and employing effective interactions
which differ in the isovector channel, 
interesting features of the E1 nuclear response were evidenced.
An essential point of our analysis is the examination of both IS and IV response
of the systems under study. 
Within our microscopic transport approach, a low-energy dipole collective mode occurs in the IV response 
of all investigated systems.  The inspection of the IS response
in the same energy region
reveals that the corresponding excitations are essentially isoscalar-like, i.e., neutrons and protons
oscillate in phase but with different amplitude. This mechanism induces a finite, though small, isovector dipole moment oscillation, which is indeed revealed in the IV strength.
These results are in agreement with the conclusions drawn from previous semi-classical investigations \cite{urbPRC2012} or from RPA studies 
\cite{mazPRC2012,Virgil2015}. 
It is worth noticing that our analysis also indicates that, in neutron-rich systems, 
the modes which are mostly isovector (such as the IV GDR) also have a
mixed character, thus contributing to the IS strength. 
Moreover, the mixing of isoscalar and isovector excitations in neutron-rich
systems has been widely discussed in the context of infinite 
nuclear matter \cite{barPR2005}. 

We also investigate how these features depend on the properties of the
effective interaction considered and, in particular, on the density behavior
of the symmetry energy.  
We observe that the strength associated with the collective pygmy dipole depends
on the symmetry energy slope. 
The analysis of the corresponding transition densities reveals that
this can be mostly related to the fact that the neutron/proton 
asymmetry of the nucleus increases, with $L$, 
at the surface, causing a larger mixing of isoscalar and isovector 
modes, which, in turn,  increases the strength observed in the isovector response.  One also observes that the asymmetry decreases, with $L$,  
in the internal part, also contributing to the dipole strength.  
Thus
the neutrons which belong to the skin play an essential
role in shaping the E1 response in the PDR region. 
However, this does not correspond to the oversimplified picture
of the PDR, associated with the oscillations of the excess neutrons against an inert isospin symmetric core. Indeed, within our transport model,  
the dynamical simulations show a more complex structure 
of the modes contributing to the PDR \cite{barPRC2012}, 
which also involves an excitation of the core, in such a way that, inside the
whole nucleus,  neutrons
and protons move in phase but with different amplitudes.  
It is also worth noticing  that these low-lying isoscalar modes are observed also in symmetric systems, 
without a corresponding IV strength in this case \cite{urbPRC2012}.

By comparing with the results obtained with simpler MI interactions,
we observe that the SAMi-J Skyrme parametrizations give 
a better reproduction of the centroid energy of the IV GDR, quite close
to the experimental value. The results of our semi-classical approach are also 
quite close to RPA calculations \cite{mazPRC2012}. 
On the other hand, the energy of the PDR looks overestimated, probably due to the semi-classical
treatment of surface effects in our approach. 

We consider that the findings presented here, in particular the connection
observed between the PDR strength, the mixed isoscalar/isovector character
of the nuclear excitations and the nuclear density profile,  can be useful for
further, systematic experiments searching for this quite elusive mode.
In particular, the features emerging from the analysis of the transition
densities may help to select the best experimental conditions to probe
the nuclear response in the PDR region. 
Moreover, a precise estimate of the strength acquired by the PDR in the dipole response
can provide indications about the neutron skin extension, 
helping to constrain yet unknown properties of the nuclear effective interaction, namely the density dependence of the symmetry energy.

\section{Acknowledgments}

We warmly thank X. Roca-Maza and G. Col\`{o} for enlightening discussions. 

This work for V. Baran was supported by a grant of the Romanian National
Authority for Scientific Research, CNCS - UEFISCDI, project number PN-II-ID-PCE-2011-3-0972.

\end{document}